\documentclass[pdflatex,sn-mathphys-num]{sn-jnl}

\usepackage{graphicx}%
\usepackage[export]{adjustbox}
\usepackage{multirow}%
\usepackage{amsmath,amssymb,amsfonts}%
\usepackage{amsthm}%
\usepackage{mathrsfs}%
\usepackage[title]{appendix}%
\usepackage{xcolor}%
\usepackage{textcomp}%
\usepackage{manyfoot}%
\usepackage{booktabs}%
\usepackage{algorithm}%
\usepackage{algorithmicx}%
\usepackage{algpseudocode}%
\usepackage{listings}%
\theoremstyle{thmstyleone}%
\theoremstyle{thmstyletwo}%

\theoremstyle{thmstylethree}%

\begin{document}

\title{Accurate Modelling of Intrabeam Scattering and its Impact on Photoinjectors for Free-Electron Lasers}

\author*[1]{\fnm{Thomas Geoffrey} \sur{Lucas}}\email{thomas.lucas@psi.ch}
\author[1]{\fnm{Paolo} \sur{Craievich}}\email{paolo.craievich@psi.ch}
\author[1]{\fnm{Eduard} \sur{Prat}}\email{eduard.prat@psi.ch}
\author[1]{\fnm{Sven} \sur{Reiche}}\email{sven.reiche@psi.ch}
\author[2]{\fnm{Erion} \sur{Gjonaj}}\email{gjonaj@temf.tu-darmstadt.de}

\affil[1]{Paul Scherrer Institut, Forschungsstrasse 111 5232 Villigen, Switzerland}
\affil[2]{Institute for Accelerator Science and Electromagnetic Fields (TEMF), Darmstadt, Germany}

\abstract{
Intrabeam scattering (IBS) is a fundamental effect that can limit the performance of high-brightness electron machines, yet it has so far been neglected in standard modelling of RF photoinjectors. Recent measurements at SwissFEL reveal that the slice energy spread (SES) in the injector is significantly underestimated by conventional tracking codes. In this work, we present a dedicated Monte Carlo simulation model that accurately predicts the IBS-induced SES growth in the photoinjector of an X-ray free-electron laser. The simulations are benchmarked against SES measurements at the SwissFEL as well as theoretically supported by a new analytical model. The results demonstrate that IBS-induced SES growth occurs throughout the injector, most prominently in the electron source, and must be taken into account when assessing photoinjector performance. We further show that while the 5D brightness is largely conserved, the 6D brightness undergoes notable degradation with propagation, underscoring the need to include IBS in the accurate design and optimization of photoinjectors.}


\maketitle


\section{Introduction}
High-brightness electron beams are critical for the lasing performance of X-ray Free-Electron Lasers (XFELs). The brightness of the beam is defined as the product of peak current and inverse phase-space volume. According to Liouville’s theorem, the 6D brightness at the cathode sets the theoretical upper limit downstream, making its preservation throughout the injector and linac a primary objective in the design and operation of XFELs.

At SwissFEL, measurements of the slice energy spread (SES) in the injector~\cite{Prat2020,Prat2022} revealed values significantly higher than predictions from standard tracking codes such as ASTRA~\cite{ASTRA_cite}, GPT~\cite{GPT_cite}, and OPAL~\cite{OPAL_cite}. Similar discrepancies have been reported at other facilities, including the European XFEL and PITZ~\cite{Tomin2021,Qian2022}. These discrepancies have been attributed to physical effects not captured by conventional macro-particle based space-charge models, including the microbunching instability (MBI) and intrabeam scattering (IBS).

The SES of the electron beam is critical for XFEL operation, as the relative SES must remain below the Pierce parameter ($\rho \sim 10^{-4} - 10^{-3}$ for X-rays) for effective lasing~\cite{Bonifacio1984}. Furthermore, a higher SES of the beam: limits the minimum pulse length, restricts the achievable wavelengths, and demands greater seed power in seeded FEL schemes. Consequently, accurate SES modelling along the injector and linac is essential to accurately determine machine performance~\cite{Prat2022}.

In this paper, we present the implementation of a Monte Carlo-based IBS model in the particle tracking code REPTIL that has been validate against measurement~\cite{Gjonaj2022}. We also present a matching analytical model that can be used for optimisation studies. These tools are used to study SES and 6D brightness evolution along the injector, emphasising the electron source. Our results highlight the importance of including IBS in the design of next-generation high-brightness XFEL electron sources and injectors as it will be seen that the concept of developing an electron source strictly around the concept of high 5D brightness may not necessarily lead to an overall increase in performance, contrary to commonly used methods.

\section{Role of IBS and 6D brightness for XFEL Performance}
\label{Subsec:6DBrightness}
The performance of an XFEL is characterised by the FEL (or Pierce) parameter, $\rho$, which in 1D-FEL theory is given by~\cite{Pellegrini2016}
\begin{equation}
       \rho := \frac{1}{\gamma}\bigg[\bigg (\frac{K f_c}{4 k_u \sigma_x} \bigg )^2 \frac{I_p}{2 I_A} \bigg]^{1/3},
       \label{Eqn:rho_definition}
\end{equation}
where $\gamma$ is the Lorentz factor, $K$ the undulator parameter, $k_u$ the undulator wavenumber, $f_c\approx1$ the coupling factor, $I_p$ the peak current, $I_A$ the Alfvén current, and $\sigma_x$ the rms beam size. Assuming cylindrical symmetry and defining $B_{5D} := \frac{2I_p}{\epsilon_N^2}$, with $\epsilon_N$ the normalised transverse emittance, we find~\cite{Pellegrini2016}
\begin{equation}
        \rho \propto B_{5D}^{1/3}.
       \label{Eqn:rho_propto}
\end{equation}

While this 1D theory highlights key dependencies, it neglects critical effects such as beam diffraction and energy spread. In particular, Landau damping reduces FEL gain when $\frac{\sigma_\gamma}{\gamma} > \rho$, with $\sigma_\gamma$ the SES. In the limiting case $\frac{\sigma_\gamma}{\gamma} \approx \rho$,  Eq.~(\ref{Eqn:rho_definition}) becomes
\begin{equation}
       \rho^2 \frac{\sigma_\gamma}{\gamma} \approx \frac{1}{I_A \gamma^3} \bigg (\frac{K f_c}{4 \sqrt{2} k_u} \bigg )^2 \frac{I_p}{\sigma_x^2}.
       \label{Eqn:rho_redefine_1}
\end{equation}
Using $\sigma_x = \sqrt{\beta \epsilon_N/\gamma}$ where $\beta$ is the optical beta function, this can be rewritten as
\begin{equation}
       \rho^2 \frac{\sigma_\gamma}{\gamma} \approx \frac{1}{I_A} \bigg (\frac{K f_c}{4 \sqrt{2} \gamma k_u} \bigg )^2 \frac{I_p}{\epsilon_N \beta}.
       \label{Eqn:rho_redefine_2}
\end{equation}
Furthermore, the FEL performance is affected by emittance-induced phase oscillations, as well as by the SES of the beam. In this limiting case, within a strong focusing undulator, the FEL parameter can be approximated by~\cite{Huang2007}
\begin{equation}
\rho \approx \frac{\epsilon_N \lambda_u}{4 \lambda \gamma \beta},
\label{Eqn:focusing_tradeoff}
\end{equation}
where $\lambda = \frac{\lambda_u}{2 \gamma^2} (1+K^2/2)$ is the resonant wavelength and $\lambda_u$ is the undulator wavelength. Substituting  Eq.~(\ref{Eqn:focusing_tradeoff}) into  Eq.~(\ref{Eqn:rho_redefine_2}), and defining the 6D brightness as $B_{6D} := \frac{2I_p}{\epsilon_N^2 \sigma_\gamma}$, we find
\begin{equation}
       \rho \propto B_{6D}.
       \label{Eqn:rho_redefined}
\end{equation}

Thus, the FEL gain is ultimately proportional to the full 6D-brightness. This highlights the importance of accurately modelling the longitudinal beam dynamics, including effects such as IBS, which give rise to SES-growth along the accelerator line.

\begin{figure*}[!htb]
    \centering
    \includegraphics[width=\linewidth]{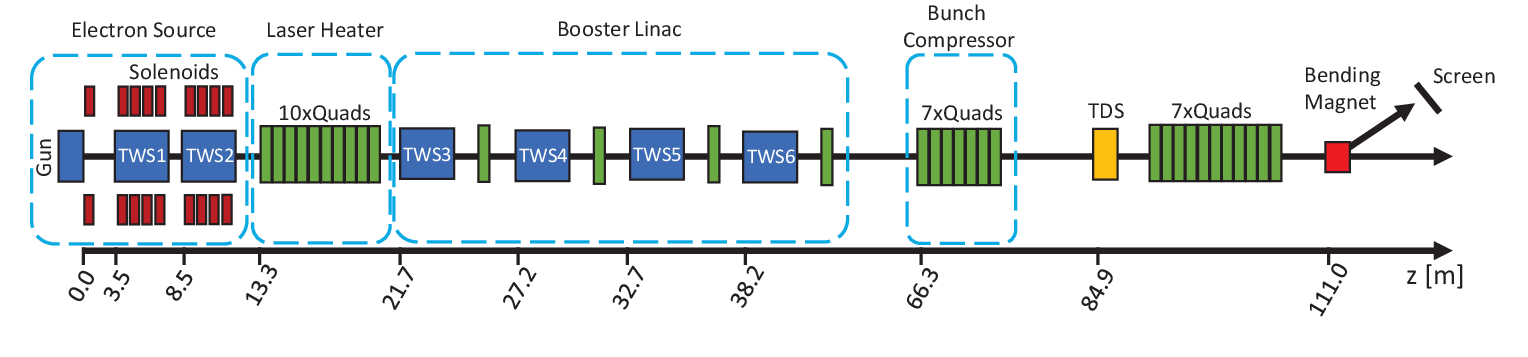}
    \caption{Layout of the SwissFEL injector~\cite{Prat2022,Prat2020_2} from the SES measurements which is used in numerical simulations. For the measurements of SES at the SwissFEL, described within this work, the R$_{56}$ of the laser heater and that of the bunch compressor are set to zero.}
    \label{Fig:Fig6}
\end{figure*}

\section{Modelling of IBS in the SwissFEL Injector}
\label{Sec:Measurements_vs_numerical}

Conventional macro-particle-based space-charge codes do not include IBS, as particle collisions occur on time-scales much shorter than the collective dynamics typically modelled. This section outlines our approach to incorporating IBS in beam dynamics simulations, which are benchmarked against SES measurements in the SwissFEL photoinjector (Fig.~\ref{Fig:Fig6}). Furthermore, we revisit the analytical treatment described in~\cite{Prat2022}.

\subsection{Numerical Simulation Model}
\label{Sec:IBS_model}

\indent To simulate collisional effects, we implemented a stochastic Monte Carlo model in the REPTIL tracking code, used here specifically for IBS modelling in XFEL injectors. The technique has been thoroughly described in~\cite{Gjonaj2022}. The model is based on Nanbu's cumulative binary collision method~\cite{Nanbu1994}, where particles are grouped into spatial cells, and random pairs within each cell undergo effective Coulomb collisions. These collisions are ultimately modelled as momenta rotations in the centre-of-mass frame of the particle pair.

The scattering angle distribution of the binary collision process is determined by the local Coulomb logarithm, which REPTIL computes dynamically using the evolving charge density and momenta distributions~\cite{Gjonaj2022}. This enables accurate IBS modelling for arbitrary distributions. This is particularly important in photoinjector simulations, where the beam distribution along the line can vary significantly.

\begin{figure}
    \centering
    \includegraphics[width=0.5\linewidth]{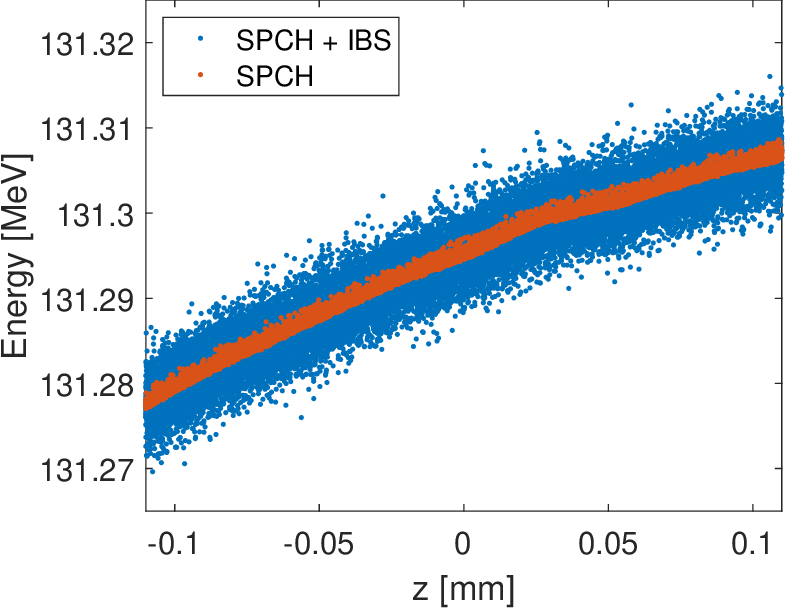}
    \caption{Detail of the longitudinal phase space of the bunch at z = 13 m with (blue) and without (red) IBS included in the space-charge (SPCH) simulations and each performed using the REPTIL code.}
    \label{fig:fig5}
\end{figure}

Fig.~\ref{fig:fig5} illustrates the impact of IBS by comparing longitudinal phase space at $z=13$~m for space-charge (SPCH) simulations with and without the IBS module. IBS introduces notable energy diffusion, broadening the local energy spread while preserving the overall energy curvature.

Previous IBS simulations with REPTIL have been reported for a simplified European XFEL injector~\cite{Gjonaj2022}. Here, we present results for the SwissFEL case and compare them to SES measurements from~\cite{Prat2022}.

\subsection{Numerical Simulations vs. Measurements}
\label{Sec:Numerical_vs_measurements}
\begin{figure}[!htb]
    \centering
    \includegraphics[width=0.5 \linewidth]{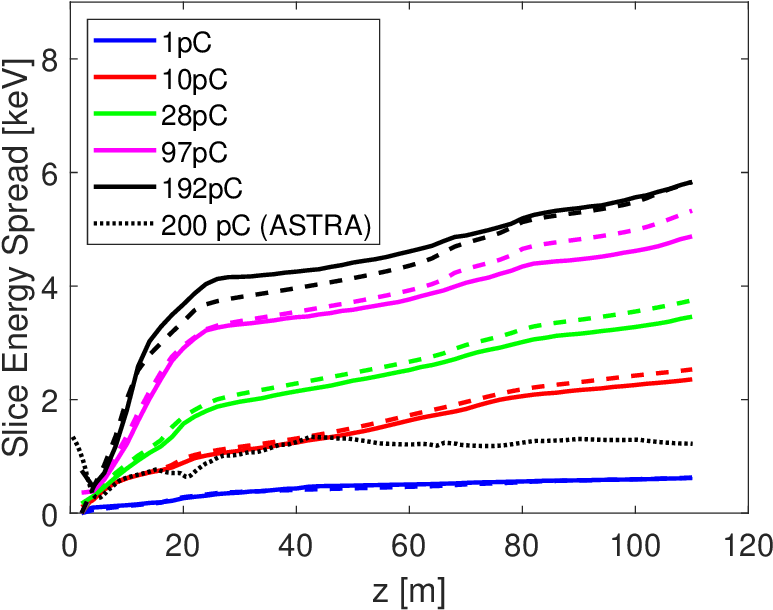}
    \caption{Slice energy spread over the SwissFEL injector, modelled in REPTIL including the IBS module (solid line) for the five different bunch charges used in the measurements~\cite{Prat2022}. The SES induced by IBS is also calculated using Eq.~(\ref{Eqn:Piwinski2}) and shown in the plot for comparison (dashed line). The SwissFEL baseline (200 pC) in ASTRA's conventional space-charge code (without IBS) is included for comparison.}
    \label{Fig:Fig7}
\end{figure}

Fig.~\ref{Fig:Fig7} shows the simulated SES along the injector using REPTIL’s IBS model for different bunch charges, using the same bunch charges as in the measurements presented in~\cite{Prat2022}. The simulations includes the full lattice from the gun to the SES measurement screen. Both the laser heater and bunch compressor chicanes had their $R_{56}$ set to zero, to prevent contributions to the SES from MBI (see e.g. the discussion in~\cite{Prat2022} on the impact of MBI). For the measurements and simulations, the laser spot size and duration on the cathode remain fixed, with only the laser intensity varied to control the bunch charge. The distribution of the laser was uniform in the transverse plane and Gaussian in the longitudinal plane.

The most notable feature is the strong SES growth in the electron source region ($z<13$~m), which slows down only when the beam enters the booster linac. This is explained by the growth of the transverse beam size, $\sigma_{x,y}$, which occurs at the end of the electron source.

Furthermore, Fig.~\ref{Fig:Fig7} also presents the slice energy spread (SES) calculated using conventional space-charge simulations in ASTRA at the nominal SwissFEL baseline charge of 200~pC. The highest measured SES of 5.8–6~keV at 192~pC is approximately an order of magnitude larger than that predicted by conventional space-charge simulations~\cite{Prat2020}.

This systematic underestimation of the SES in conventional space-charge codes results in a predicted 6D brightness that is nearly an order of magnitude higher than that obtained in experimental measurements. Consequently, accurate modelling of IBS is essential for a realistic assessment of the XFEL performance.

\begin{figure}[!htb]
    \centering
    \includegraphics[width=0.5 \linewidth]{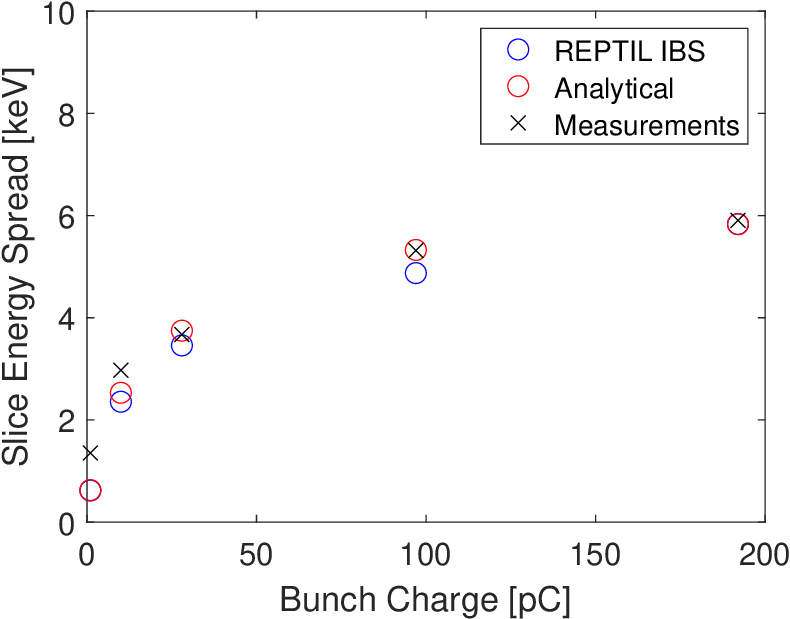}
    \caption{Comparison between the measurements at the SwissFEL, the numerical simulations with IBS performed with REPTIL and the analytical calculation using Eq.~(\ref{Eqn:Piwinski2}). The values are given 111~m downstream of the cathode.}
    \label{Fig:Fig8}
\end{figure}

Figure~\ref{Fig:Fig8} compares the simulated SES at the injector exit (111~m) with experimental measurements across various bunch charges. The agreement between the numerical simulations and measurements is strong, with both exhibiting a square-root dependence on bunch charge, in line with the Piwinski model~\cite{Piwinski1974}.

\subsection{Analytical Model}
In circular machines, analytical models are often used to complement numerical simulations. This is because Monte Carlo-based IBS simulations are computationally intensive, requiring at least 500k macro-particles per bunch for numerical convergence. This limits their practicality in optimisation studies. Below, we introduce a modified analytical model based on Piwinski's IBS-model that is applicable in the case of a low-energy injectors~\cite{Huang2002}.

In~\cite{Huang2002}, Huang reformulated Piwinski’s model for linacs by neglecting synchrotron oscillations. The SES growth for a 3D Gaussian bunch over a distance $\Delta z$ is given by:
\begin{equation}
    \sigma_\gamma^2 = \sigma_{\gamma,0}^2 + \frac{r_e N_b \Lambda_c}{4 \sigma_x \epsilon_N \sigma_z} \Delta z,
    \label{Eqn:Piwinski}
\end{equation}
where $\sigma_{\gamma,0}$ is the initial SES, $r_e$ the classical electron radius, $N_b$ the number of electrons, $\sigma_z$ the rms bunch length, and $\Lambda_c$ the Coulomb logarithm.

To account for the local nature of IBS, we will use the mean slice size over x and y ($\sigma_{r,s}=\sqrt{\sigma_{x,s}\sigma_{y,s}}$) and the slice normalized emittance ($\epsilon_{N,s}$) of the {\em slice} instead of the corresponding rms-valued quantities used in~\cite{Huang2002}. Since IBS in our case is dominated by particle collisions in transverse planes (whereas the longitudinal space-charge interaction is negligible), this slice-based treatment is more appropriate. In particular, our measurements mention the SES of the {\em central slice} therefore we use the parameters of this particular slice.
We can also relate $N_b/\sigma_z$ to the peak current as 
\begin{equation}
    N_b/\sigma_z = \sqrt{2 \pi} \frac{q_s}{c e \Delta z} = \frac{\sqrt{2 \pi} I_p}{c e}
    \label{Eqn:peak_current}
\end{equation}
where $q_s$ is the charge in the central slice, $\Delta z_s$ is the full slice length and $I_p$ is the peak current. In the above, we have modelled the central slice as a Gaussian beamlet of rms length $\sigma_{z,s}=\Delta z/\sqrt{2\pi}$. This allows to apply directly Piwinski's theory for Gaussian distributions in  Eq.~(\ref{Eqn:Piwinski}) for the IBS growth rate within a single slice of the bunch.
Substituting into  Eq.~(\ref{Eqn:peak_current}) into Eq.~(\ref{Eqn:Piwinski}) yields:
\begin{equation}
    \sigma_\gamma^2 = \sigma_{\gamma,0}^2 + \frac{\sqrt{2 \pi} r_e I_p \Lambda_c}{4 c e \sigma_{r,s} \epsilon_{N,s}} \Delta z.
    \label{Eqn:Piwinski2}
\end{equation}
The Coulomb logarithm is evaluated locally for the slice following the argument in~\cite{Huang2002} as:
\begin{equation}
    \Lambda_c = \ln\left(\frac{\Delta \gamma_{\max}}{\Delta \gamma_{\min}}\right), \quad 
    \Delta \gamma_{\max} = \gamma^2 \sigma_{r',s}, \quad
    \Delta \gamma_{\min} = \frac{r_e}{\sigma_{r,s} \sigma_{r',s}}.
\end{equation}
where $\sigma_{r',s}$ is the mean slice beam divergence.

\subsection{Analytical Model vs. Numerical Model and Measurements}
Using Eq.~(\ref{Eqn:Piwinski2}) and the slice parameters from particle tracking simulations, we can compare to the numerical calculations and measurements. The slices are defined by the longitudinal distribution, of which we take the longitudinally central slice as done in the measurements. The number of slices was chosen such that the total energy spread of the central slice converged with an increasing number of slices.

We observe strong agreement with both the measurements and numerical IBS simulations (Figs.~\ref{Fig:Fig7} and~\ref{Fig:Fig8}). This supports the validity of the analytical model. Notably, the $\sqrt{2 \pi}$ factor naturally recovers the empirical correction factor of 2.4 used in~\cite{Prat2022}.

\begin{figure}[!htb]
\centering
\begin{tabular}{ccc}
\rotatebox{90}{} 
\includegraphics[width=.3\linewidth,valign=m]{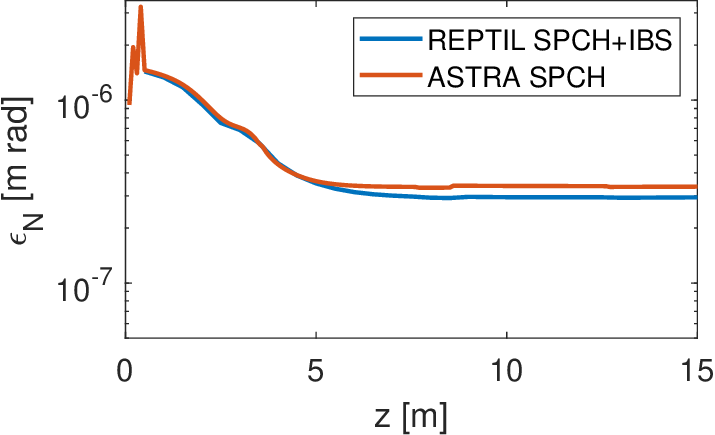} & 
\includegraphics[width=.3\linewidth,valign=m]{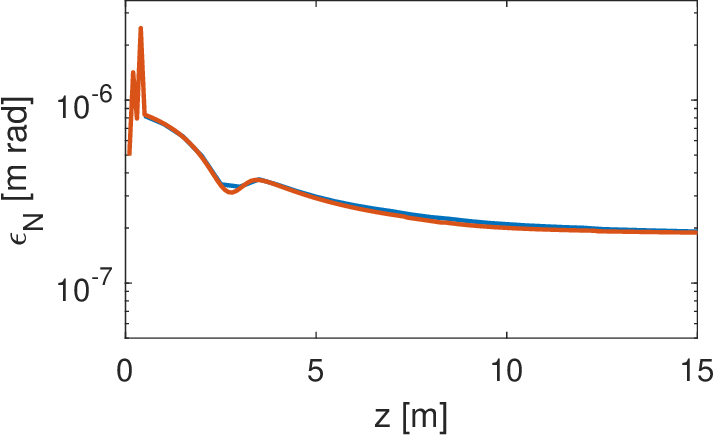} & \includegraphics[width=.3\linewidth,valign=m]{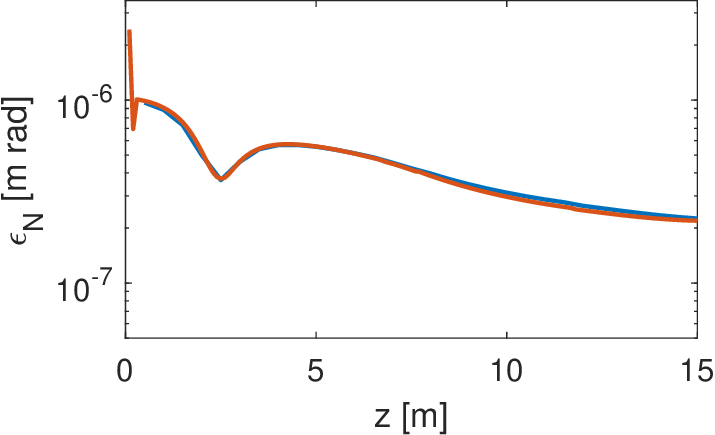}\\
\includegraphics[width=.3\linewidth,valign=m]{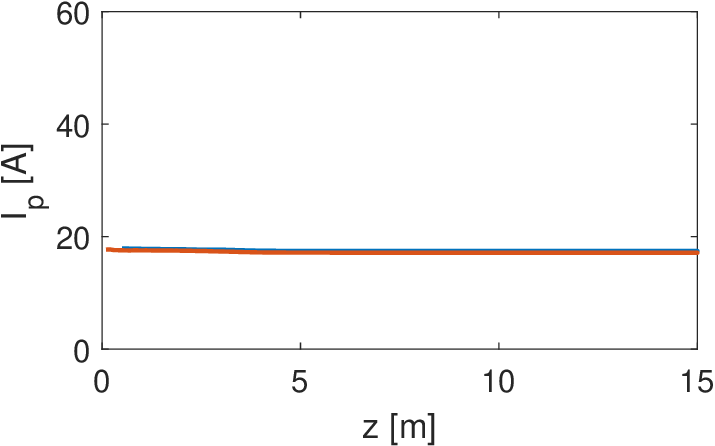} & 
\includegraphics[width=.3\linewidth,valign=m]{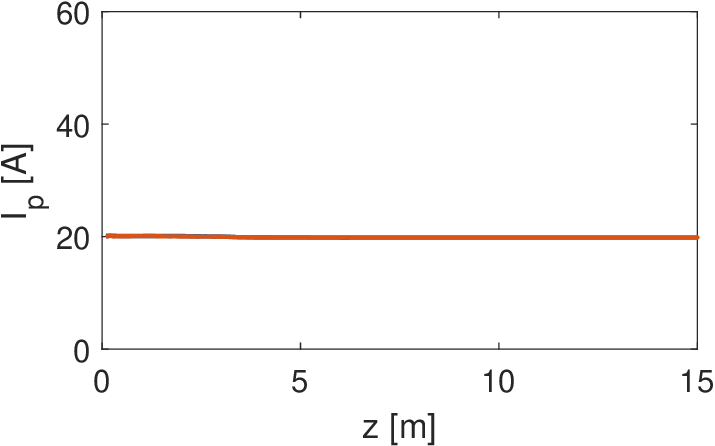} & \includegraphics[width=.3\linewidth,valign=m]{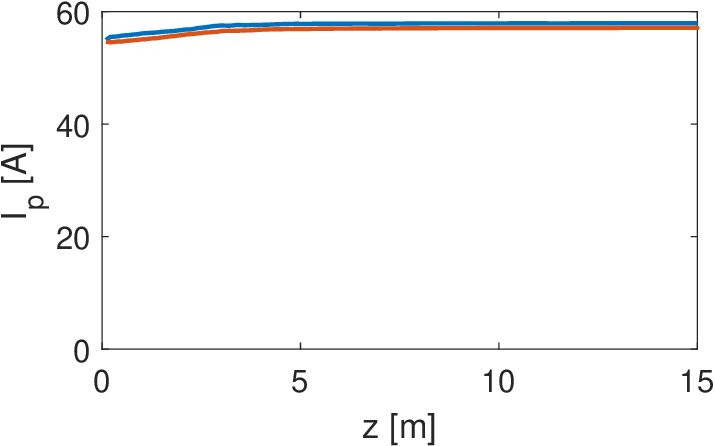}\\
\includegraphics[width=.3\linewidth,valign=m]{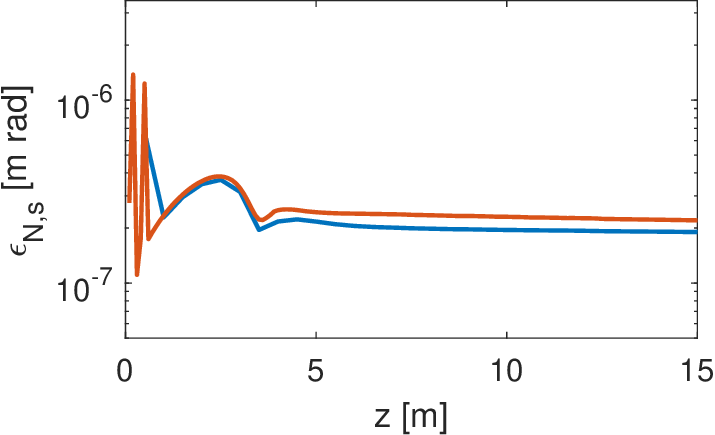} & 
\includegraphics[width=.3\linewidth,valign=m]{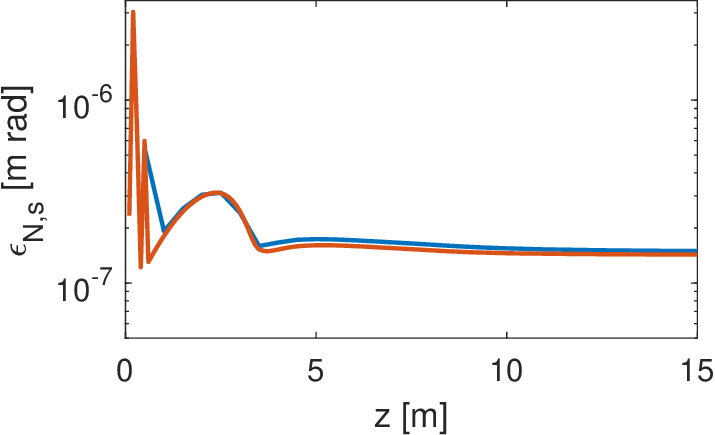} & \includegraphics[width=.3\linewidth,valign=m]{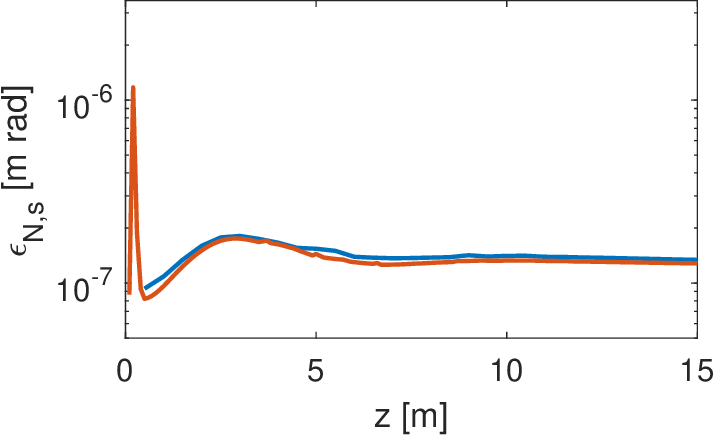}\\
\includegraphics[width=.3\linewidth,valign=m]{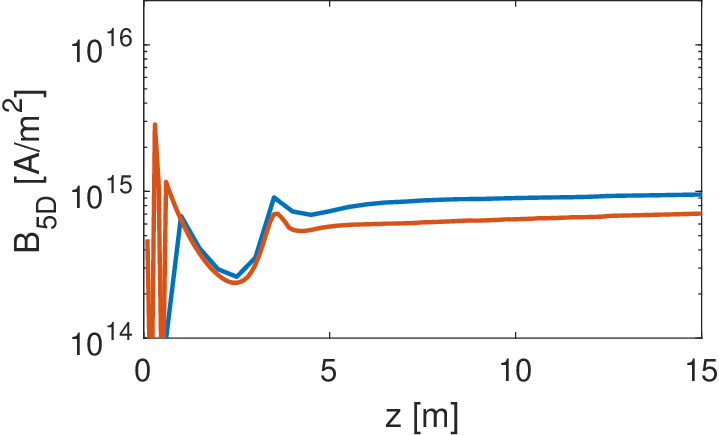} & 
\includegraphics[width=.3\linewidth,valign=m]{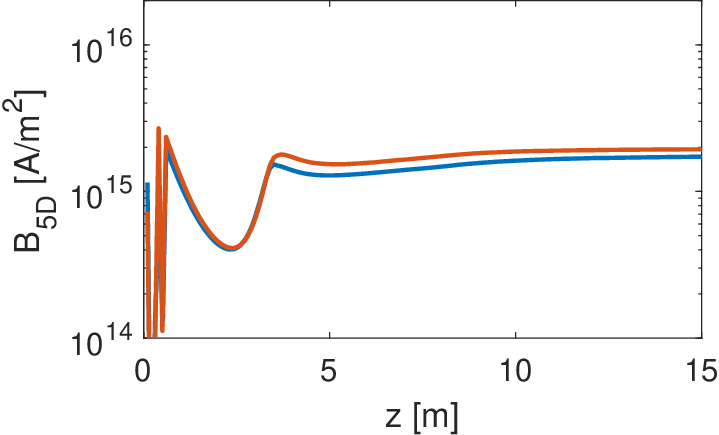} & \includegraphics[width=.3\linewidth,valign=m]{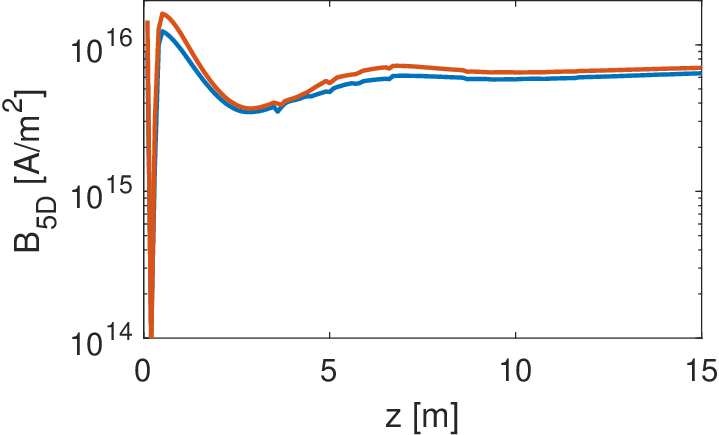}\\
\includegraphics[width=.3\linewidth,valign=m]{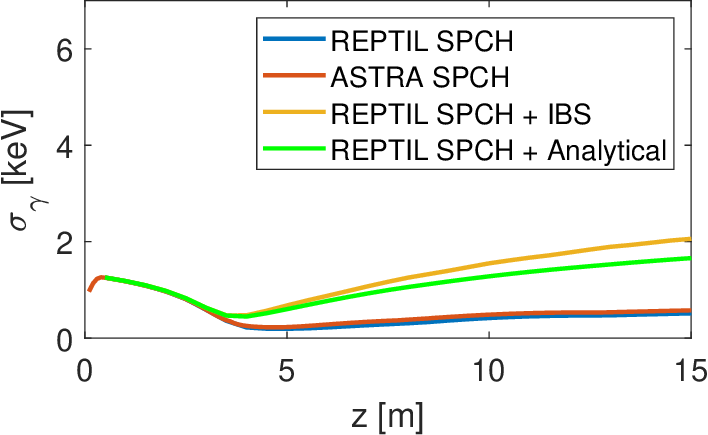} & 
\includegraphics[width=.3\linewidth,valign=m]{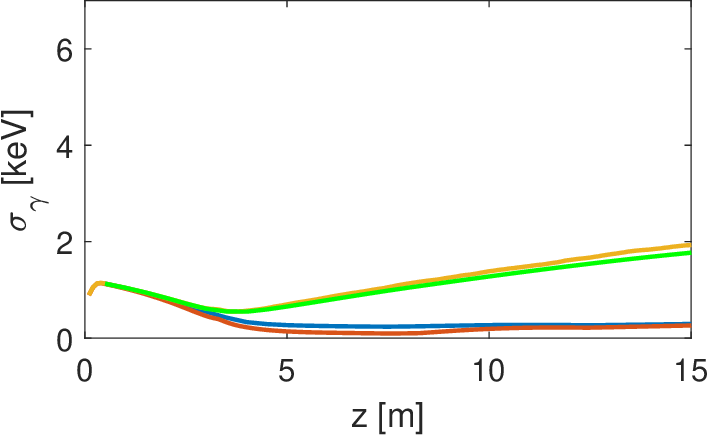} & \includegraphics[width=.3\linewidth,valign=m]{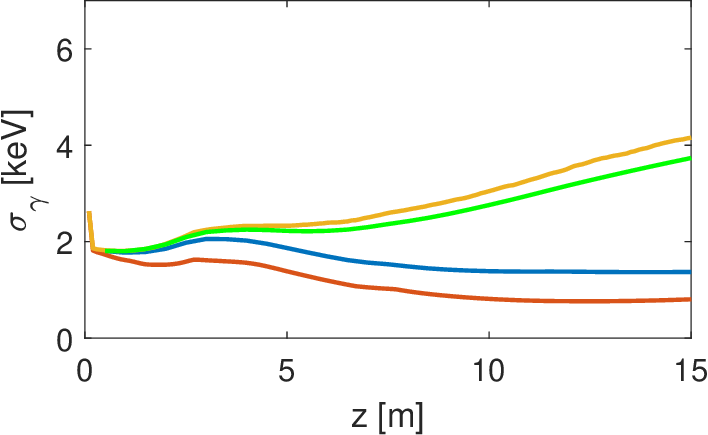}\\
\includegraphics[width=.3\linewidth,valign=m]{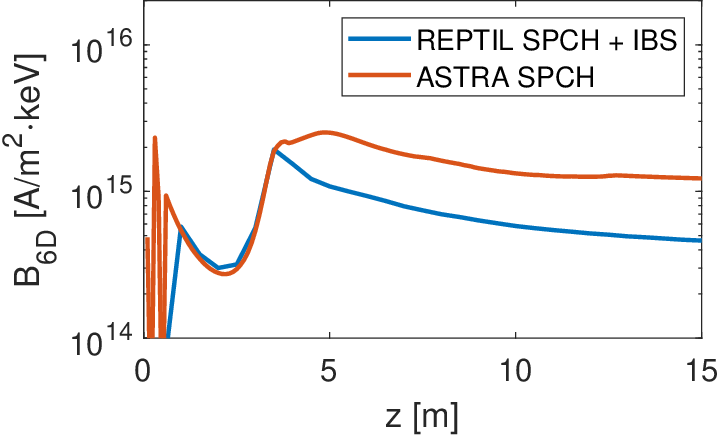} & 
\includegraphics[width=.3\linewidth,valign=m]{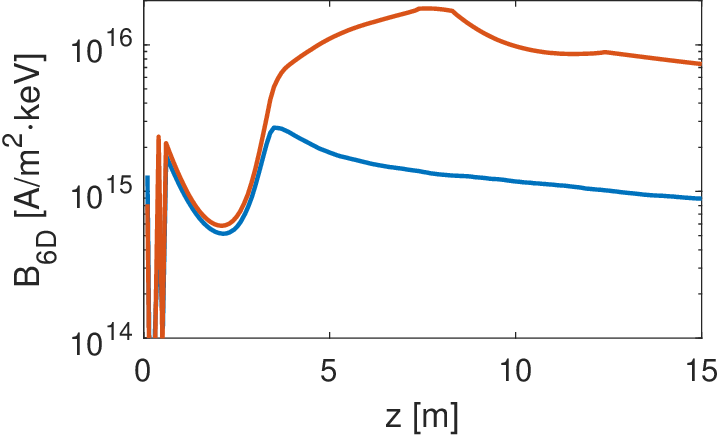} & \includegraphics[width=.3\linewidth,valign=m]{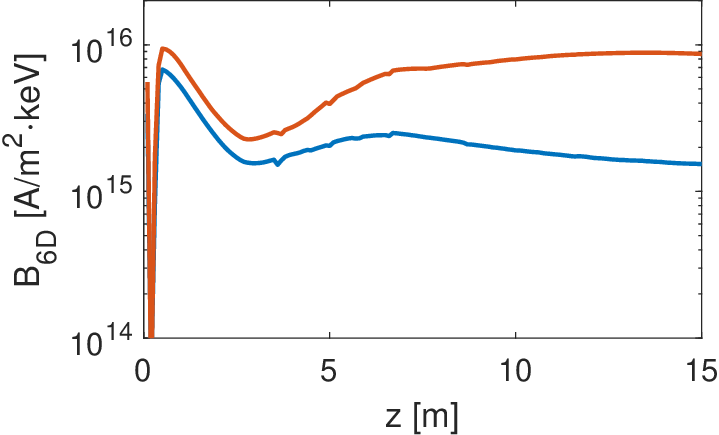}\\
\end{tabular}
\caption{Evolution of the projected emittance, peak current, central slice emittance, 5D brightness, SES and 6D brightness (top to bottom) calculated for the SwissFEL RF photogun with Gaussian longitudinal distribution (left), SwissFEL RF photogun with uniform longitudinal distribution (middle) and a proposed higher brightness travelling-wave photogun with uniform distribution (right). The project emittance, peak current and slice emittance are simulated with REPTIL and bench-marked against ASTRA. The SES was calculated with a conventional space-charge only simulation (SPCH), using the IBS module of REPTIL (SPCH $+$ IBS) and using the analytical model in Eq.~(\ref{Eqn:Piwinski2}) (SPCH only with analytical IBS).}
\label{Fig:Fig4}
\end{figure}

\section{Impact of IBS on the Brightness of Electron Sources}
\label{Sec:ElectronSourceUpgrade}
As shown in Eq.~(\ref{Eqn:Piwinski2}), increasing peak current or reducing slice emittance, hallmarks of high 5D brightness beams, also amplifies IBS-induced SES. This raises concerns about evaluating new electron sources on the basis of their 5D brightness, which is often used in the literature as the figure-of-merit. In~\cite{Lucas2023}, a new high-brightness electron source was proposed to generate a beam of higher 5D brightness from the cathode. Here, we assess the impact of IBS on this design in comparison with the existing SwissFEL RF photogun. While our study focuses on these two electron sources, IBS effects are expected to be relevant for all high-brightness electron sources and photoinjectors.  

We perform numerical tracking simulations using the REPTIL code and benchmark these against ASTRA, which is a widely validated tool. Three scenarios are considered: (a) the SwissFEL gun with Gaussian longitudinal distribution for the photocathode laser, (b) the SwissFEL gun with a uniform longitudinal distribution  for the photocathode laser, and (c) the proposed traveling-wave (TW) gun with a uniform longitudinal distribution for the photocathode laser. This allows us to examine the role of the initial longitudinal distribution of the electron bunch (Gaussian vs. uniform) on IBS. Furthermore, we can assess how changes in beam brightness impact IBS and to also validate the analytical model for different initial distributions. The latter is important as it is common to optimise an electron source with the idealised uniform distribution although such a longitudinal distribution is difficult to achieve practically.

For these three cases, REPTIL simulations are run with the standard space-charge (SPCH) model and with Space-charge with IBS enabled (SPCH+IBS). All simulations include the electron source up to 15~m, as shown in Fig.~\ref{Fig:Fig6}, to capture beam evolution along $z$ for the electron source component of the injector.

Fig.~\ref{Fig:Fig4} compares the central slice parameters, as well as the projected emittance, for the three cases. The slice emittance and peak current remain nearly constant along $z$, indicating that these beam qualities are established early in the electron source and preserved by rapid acceleration. This is consistent with the nearly constant 5D brightness derived from these parameters, observed downstream of the point where slice-mixing subsides (approximately 5~m downstream of the cathode). The effect of IBS on the emittance and peak current was found to be negligible compared to the effects of space-charge and within the resolution of the plots, so only the SPCH+IBS curve of REPTIL is shown.

Fig.~\ref{Fig:Fig4} also shows the SES evolution for both electron sources using standard SPCH simulations (blue), SPCH+IBS simulations (yellow), the analytical model of  Eq.~(\ref{Eqn:Piwinski2}) applied to slice parameters from the SPCH runs (green) and the ASTRA simulation as a reference (red). We observe a strong growth in the SES at the start of the first accelerating structure (approximately 3~m) when including IBS, which is not found using conventional space-charge modelling. Furthermore, we can use the SES to understand the machine performance that above was illustrated to scale with the 6D brightness. Unlike the 5D brightness, the 6D brightness decreases along $z$ due to IBS-driven SES growth leading to a deterioration of beam quality along $z$. This is observed in all electron source cases. When comparing the S-band SW and C-band TW electron guns for 6D brightness, we see that the TW gun continues to have a greater brightness than the existing SwissFEL gun. However, performance improvement for this new electron source is less than originally anticipated based on the estimation of the 5D brightness~\cite{Lucas2023}. This effect is due to the presence of IBS, which so far had been omitted in the analysis.

Finally, we investigate the role of the initial bunch distribution on IBS-induced SES. Monte Carlo simulations for both Gaussian and uniform distributions agree well with the analytical model, showing that the choice of initial laser distribution has little impact on the analytical calculation compared to slice parameters such as peak current, bunch size, and emittance. However, for precise evaluation of distribution-dependent effects, Monte Carlo--based IBS calculations remain preferable.

\section{Conclusions}
\indent Recent SES measurements at SwissFEL confirm that IBS plays a key role in limiting 6D beam brightness. This effect is typically not included in conventional beam dynamics simulations. For the analysis in this paper, we apply first-principle beam dynamics simulations including particle collision effects. In addition, we develop an accurate analytical IBS model, based on Piwinski's theory, which is better suited for electron injector calculations. In both cases, we obtain excellent agreement with the measurement data for the SwissFEL injector reported in~\cite{Prat2022}.

IBS has the greatest impact early in the injector, especially in the electron source. Simulations of both the current SwissFEL electron source and a prospective upgrade show that while 5D brightness is largely conserved along $z$ at the injector, 6D brightness decreases due to IBS-driven SES growth. This underscores the need to account for IBS effects in electron source design, particularly in the case that performance is strongly SES-dependent, such as seeded-FELs or where strong compression is required.

With our improved understanding of IBS, the next step is to fully model the SwissFEL injector up to the bunch compressor, incorporating our proposed electron source upgrades and accounting for IBS effects. At the bunch compressor we expect the SES to increase linearly with the compression factor therefore accurate modelling of the IBS is key.

\end{document}